
\input phyzzx
\def\ucla{Department of Physics\break
      University of California Los Angeles\break
        Los Angeles, California 90024--1547}
\def\nl{\hfil\break}

\def\ijp{{ \sl Int. J. of Mod. Phys. }}

\def\mpl{{ \sl Mod. Phys. Lett. }}

\def\npb{{ \sl Nucl. Phys. }}

\def\prd{{ \sl Phys. Rev. }}
\def\prl{{ \sl Phys. Rev. Lett. }}
\def\plb{{ \sl Phys. Lett. }}
\def\rmp{{ \sl Rev. Mod. Phys. }}

\def\undertext#1{\vtop{\hbox{#1}\kern 1pt \hrule}}
\def\half{{1\over2}}
\def\c#1{{\cal{#1}}}
\def\slash#1{\hbox{{$#1$}\kern-0.5em\raise-0.1ex\hbox{/}}}
\font\cmss=cmss10 \font\cmsss=cmss10 at 7pt
\def\IZ{\relax\ifmmode\mathchoice
{\hbox{\cmss Z\kern-.4em Z}}{\hbox{\cmss Z\kern-.4em Z}}
{\lower.9pt\hbox{\cmsss Z\kern-.4em Z}}
{\lower1.2pt\hbox{\cmsss Z\kern-.4em Z}}\else{\cmss Z\kern-.4em Z}\fi}
\overfullrule=0pt

\overfullrule=0pt
\Pubnum={UCLA/91/TEP/33}
\date={August, 1991}
\titlepage
\title{Correlation functions of minimal models coupled to
two dimensional quantum supergravity
\foot{\rm Work supported in part by the National Science Foundation grant
PHY--89--15286.}}
\author{\rm Kenichiro Aoki and Eric D'Hoker\foot{\rm
email:~aoki@uclahep.bitnet, dhoker@uclahep.bitnet} }
\address{\ucla}
\abstract{
We compute the three point functions of Neveu--Schwarz
primary fields
of the minimal models on the sphere when coupled
to supergravity in two dimensions.
The results show that the three point correlation functions
are determined by the scaling dimensions of
the fields, as in the bosonic case.}
\medskip
\endpage
\def\al#1{\alpha_{#1}}
\def\alr#1{\alpha_{r'_{#1}r_{#1}}}
\def\bestar{\alpha}
\def\frac#1{\left\langle{#1}\right\rangle}

\def\corr#1{\left\langle{#1}\right\rangle}
\def\prodone#1{\prod_{j=1}^{#1}}

\def\d{\partial}
\def\inv#1{{#1}^{-1}\!}
\def\beh#1{\beta(h_{#1})}

\def\Op{\Psi}
\def\Opcft#1{\Psi^{matter}_{#1}(\zz_{#1})}
\def\QQ{\kappa}
\def\zz{{\bf z}}
\def\ww{{\bf w}}
\def\D{D}
\def\varPhi{\Phi'}
\def\rt{\rho}
\def\invt{\nabla}
\def\invtb{\overline{\invt}}
\def\inbar{\,\vrule height1.5ex width.4pt depth0pt}
\def\IC{\relax\hbox{$\inbar\kern-.3em{\rm C}$}}
\def\th#1{\theta_{z_{#1}}}
\REF\KPZ{A. M. Polyakov, \mpl {\bf A2} (1987) 893 \nl
V. Knizhnik, A. M. Polyakov, A. Zamolodchikov,
\mpl {\bf A3} (1988) 819\nl
A. M. Polyakov, A. Zamolodchikov, \mpl {\bf A3} (1988) 1213}
\REF\BDS{E. Br\'ezin, V. Kazakov, \plb {\bf B236} (1990) 144 \nl
	M. Douglas, S. Shenker, \npb {\bf B335} (1990) 635 \nl
	D. J. Gross, A. A. Migdal, \prl {\bf 64} (1990) 127 }
\REF\TOP{E. Witten, \npb {\bf B340} (1990) 280\nl
	J. Distler, \npb {\bf B342} (1990) 523\nl
	R. Dijkgraaf, E. Witten, \npb {\bf B342} (1990) 486\nl
	E. Verlinde, H. Verlinde, \npb{\bf B348} (1991) 457
}
\REF\POLYAKOV{A.M. Polyakov, \plb {\bf B103} (1981) 207, 211\nl
	T. Curtright and C. Thorn, \prl 48 (1982) 1309\nl
	E. D'Hoker and R. Jackiw, \prd D26 (1982) 3517\nl
	J.-L. Gervais and A. Neveu, \npb B209 (1982) 125}
\REF\GERVAIS{O. Babelon, \npb {\bf B258} (1985) 680\nl
E. D'Hoker, \prd{\bf D28} (1983) 1346\nl
T. Curtright, G. Ghandour, \plb{\bf 136B} (1984) 50\nl
A.~Bilal, J.-L. Gervais, \npb{\bf B293} (1987) 1}
\REF\DDK{F. David, E. Guitter, {\sl Euro. Phys. Lett.}
{\bf 3} (1987) 1169\nl
	F. David, \mpl {\bf A3} (1988) 1651\nl
	J. Distler, H. Kawai, \npb {\bf B321} (1988) 509\nl
	J. Distler, H. Kawai, Z. Hlousek, \ijp{\bf A5} (1990) 391,1093}
\REF\PURESUGY{K. Aoki, D. Montano,
J. Sonnenschein, \plb {\bf 247B} (1990) 64\nl
J. Hughes, K. Li, \plb{\bf 264B} (1991) 261\nl
H.~Yoshii, \plb{\bf 259B} (1991) 279
}
\REF\SUPERDDK{P.~Di Francesco, J.~Distler, D.~Kutasov,
\mpl {\bf A5} (1991) 2135}
\REF\GL{M. Goulian, M. Li, \prl {\bf 66} (1991) 2051\nl
P.~Di Francesco, D.~Kutasov, \plb {\bf 261B} (1991) 385\nl
Vl. S. Dotsenko, PAR-LPTHE 91-18 (1991) preprint\nl
Y. Kitazawa, HUTP-91/A013 (1991) preprint}
\REF\AD{K.~Aoki, E.~D'Hoker, /UCLA/91/TEP/32 (1991) preprint}
\REF\BPZ{A. Belavin, A. M. Polyakov, A. Zamolodchikov,
	\npb {\bf B241} (1984) 333}
\REF\SMIN{M. Bershadsky, V.G. Knizhnik, M.G. Teitelman,
\plb {\bf 151B} (1985) 31\nl
D.~Friedan, Z. Qiu, S. Shenker, \plb {\bf 151B} (1985) 37}
\REF\KITETAL{Y.~Kitazawa, N.~Ishibashi, A.~Kato, K.~Kobayashi, Y.~Matsuo,
S.~Odake, \npb {\bf B306} (198) 425}
\REF\FQS{D. Friedan, Z, Qiu, S. Shenker, \prl {\bf 52} (1984) 1575}
\REF\CIZ{A. Cappeli, C.~Itzykson, J.-B.~Zuber \npb {\bf B280} (1987) 445\nl
D.~Gepner, Z.~Qiu, \npb {\bf B285} (1987) 423}
\REF\CAPPELI{A. Cappeli, \plb {\bf 185B} (1987) 82
 \nl D.~Kastor, \npb {\bf B280} (1987) 304
}
\REF\RMP{E.~D'Hoker, D.H.~Phong, \rmp {\bf 60} (1988) 917}
\REF\DF{Vl. S. Dotsenko, V. A.  Fateev, \npb {\bf B240} (1984) 312,
	B251 (1985) 691, \plb {\bf 154B} (1985) 291}
\REF\BD{M. Staudacher, \npb {\bf B336} (1990) 349\nl
E.~Br\'ezin, M.R.~Douglas, V.~Kazakov, S.H.~Shenker, \plb{\bf237B} (1990) 43}
\REF\AG{L.~Alvarez--Gaum\'e, J.L.~Manes, CERN--TH.6067/91 (1991) preprint}
\REF\BK{M.~Bershadsky, I.~Klebanov, \prl {\bf 65} (1990) 3088}
\REF\PS{V.~Periwal, D.~Shevitz, \npb {\bf B344} (1990) 731\nl
K. Demeterfi, C.-I. Tan, \mpl {\bf A5} (1990) 1563\nl
\v C. Crnkvi\'c, M.R. Douglas, G. Moore, YUTP-P6-90,RU-90-62 (1991) preprint}
\lettersize\singlespace
\chapter{Introduction}
Currently, there are several approaches to two dimensional
quantum gravity including the  light--cone gauge approach [\KPZ],
matrix models [\BDS], topological field theory [\TOP]
and conformal matter coupled to Liouville theory
[\POLYAKOV,\GERVAIS,\DDK].
However, in the case of two dimensional supergravity,
no satisfactory matrix model formulation has been
constructed. In the topological field theory approach,
even though some progress has been made,
correlation functions are yet to be computed [\PURESUGY,\SUPERDDK].

One advantage the super Liouville approach  has over the others is that
it is straightforward, at least conceptually,
to generalize the facts established in the bosonic case to the
supersymmetric case.
The three point functions were computed in the bosonic
case using the area operator as a screening charge and then
using the results from  conformal field theory [\GL].
A crucial ingredient in their method is a
 formal continuation of  an identity shown on  the
integers to  non--integer values, which was
justified using a semi--classical analysis in [\AD].

In this work, we shall compute the genus zero three point functions
of minimal models  [\BPZ,\SMIN] coupled to two dimensional supergravity
using the Liouville theory approach [\DDK].
The computation is performed for general Neveu-Schwarz primary
fields in the $N=1$ minimal models, including the non--unitary ones.
In fact, the results may be stated simply; the three point functions
on the sphere are determined
by the scaling dimensions of the operators, as in the
bosonic case.
The Liouville approach is the most natural one from the field theoretical
point of view.
We first compute the correlation function for a fixed metric
and integrate over the metric, modulo symmetries.
Weyl invariance,  which was present in the classical theory,  is destroyed
by the conformal anomaly, so that the integration over the metric corresponds
to the integration over the moduli space of Riemann surfaces and the scale
factor of the metric using the Liouville measure at a fixed fiducial
metric.
The correlation functions for the fixed metric
are none other than the conformal
field theory correlation functions.
The conformal field theory
three point functions on the sphere for the minimal
models have been computed in
[\KITETAL] for the supersymmetric case.
Another strong point of   the Liouville approach is that the matter
sector has been actively studied and
is reasonably well understood from the works
trying to classify conformal field theory.
In particular, the so called minimal models and their supersymmetric
generalizations have been classified [\BPZ,\SMIN,\FQS,\CIZ,\CAPPELI].
 \chapter{Minimal models coupled to supergravity}
In this section, we compute the three point functions of
the minimal models coupled to two dimensional
supergravity.
The calculation parallels that of the bosonic case in [\GL,\AD].
We will work in superspace and follow the conventions
of [\RMP].
We shall formulate the super Liouville approach to two dimensional
quantum supergravity  along the lines of [\DDK].
A general correlation function on a genus $p$ surface
is given by (genus zero and one are special due to
the existence of continuous conformal automorphisms and
will be explained below.)
$$\corr{\prodone N\Op_j}
=\int_{s\c M_{p}}\!\!\!dm\int \!\!D_{\!\!\hat E}\Phi\,e^{-S_{SL}}
\prodone N \int d^2\zz_j
e^{\beta(h_j)\Phi(\zz_j)}\corr{\prodone N\Opcft j}_{\hat E}\eqno\eq$$
where $s\c M_p$ denotes the moduli space of genus $p$ super
Riemann surfaces and the super Liouville action $S_{SL}$ is
$$ S_{SL} = {1\over4\pi}\int{\hat E}
   \left[\half\D_\alpha\Phi\D^\alpha\Phi
  -\QQ\hat R\Phi+\mu e^{\bestar\Phi}
   \right],\eqno\eq$$
The parameters in the super Liouville action may be determined, by the
conditions that the matter part combined with the gravitational
part is conformal invariant and that
$\int\exp(\beta(h)\Phi)\Psi^{matter}$
should have conformal dimension $(0,0)$,
to be
$$ \QQ=\sqrt{9-\hat c\over2},\quad
\beta(h) ={-\sqrt{9-\hat c}+\sqrt{1-\hat c+16h}\over2\sqrt2}
,\quad \bestar\equiv\beta(h_{11}) . \eqno\eq$$
$\hat c$ denotes the amount of matter coupled to supergravity
measured in the units of scalar superfields.
We shall specialize to the case of $N=1$ minimal model
in the Neveu--Schwarz sector. Then
$$ \eqalign{\hat c&=1-{8\over q(q+2)},
\quad h_{r'r}={\left(qr'-(q+2)r\right)^2-4\over8q(q+2)}, \cr
\QQ&={2(q+1)\over\sqrt{q(q+2)}},
\quad\beta(h_{r'r}) ={-2(q+1)+\left|qr'-(q+2)r\right|\over2\sqrt{q(q+2)}}
,\quad\bestar=-\sqrt{q\over(q+2)} .\cr} \eqn\Dimeq$$
Integer $q$ corresponds to the unitary minimal models and $q$ is
rational for general non--unitary models.
We should point out that the method of screening charges
which we use to compute correlation functions is valid
even for irrational $q,r'_j,r_j$ at tree level as long
as the sum of the external charges is integer.
However, for generic values of $q$, it is not known
how to construct interacting theories.

Integrating over the constant mode of $\Phi$,
$$ \int D_{\!\!\hat E}\Phi\,e^{-S_{SL}}\prodone N e^{\beta(h_j)\Phi(\zz_j)} =
\left({\mu\over2\pi}\right)^s\inv{\bestar}\Gamma(-s)
\int D_{\!\!\hat E}'\Phi e^{-S'_{SL}}\left(\int{\hat E}
  e^{\bestar\Phi}\right)^s
\prodone N e^{\beta(h_j)\Phi(\zz_j)}\eqno\eq$$
where $D'_{\!\!\hat E}\Phi$ denotes the integration over the modes orthogonal
to the constant mode and
 $$ S'_{SL}\equiv{1\over4\pi}\int{\hat E}
   \left[\half\D_\alpha\Phi\D^\alpha\Phi
  -\QQ\hat R\Phi    \right],\qquad
s=-{\QQ\over\bestar}(1-p)-\sum_{j=1}^N{\beta(h_j)\over\bestar}\eqn\Seq$$

Let us now specialize to correlation functions on the sphere.
The sphere has no moduli but has a group of conformal
automorphisms that act on the surface, Osp$(2|1;\IC)$
of complex dimension $3|2$.
Therefore, we integrate over the location of the
operators, $\zz_j$, and then divide by the volume of
the group of conformal automorphisms.
This corresponds to
integrating with the measure
$|(\zz_1-\zz_2)(\zz_2-\zz_3)(\zz_3-\zz_1)|d^2\invt$
 after combining the gravitational
part with the matter part of the correlation function.
Here, $\invt,\invtb$ are invariants under Osp$(2|1;\IC)$.
\foot{
We shall use the  notations
$\zz\equiv z|\theta_z,\quad\ww\equiv w|\theta_w,\quad
\zz-\ww\equiv z-w-\theta_z\theta_w,\ \etc$. $\invt$ is defined as
$$\invt\equiv{\th1(\zz_2-\zz_3)+\th2(\zz_3-\zz_1)+\th3(\zz_1-\zz_2)+
\th1\th2\th3\over\left[\left(\zz_1-\zz_2\right)
\left(\zz_2-\zz_3\right)\left(\zz_3-\zz_1\right)\right]^{1/2}}$$}
We will concentrate the curvature of the sphere at infinity,
$\infty|0$ and work with the flat zweibein on the plane.
$$ \eqalign{\int D_{\!\!\hat E}\Phi\,e^{-S_{SL}}
\prodone 3 e^{\beta(h_j)\Phi(\zz_j)} &=
\left({\mu\over2\pi}\right)^s{\Gamma(-s)\over\bestar}
  \prod_{i<j\atop1}^3|\zz_i-\zz_j|^{-2\beh i\beh j}
\cr&\qquad\times
  \int\prod_{j=1}^{s}d^2\ww_j
  \prod_{i=1}^{3}|\zz_i-\ww_j|^{-2\bestar\beh i}
  \prod_{i<j\atop i,j=1}^{s}|\ww_i-\ww_j|^{-2\bestar^2}.\cr}
\eqno\eq$$

The general integral of this type
may be deduced from the work of [\KITETAL], possibly up
to a normalization factor for the screening charges, which
may be absorbed into the normalization of the external fields.
(Below, $\frac{x}$ denotes the fractional part of $x$, \ie\
$\frac{x}\equiv x-[x]$.)
$$\eqalign{{\cal J}_{n'n}&(a'_j,a_j,\rt',\rt;\zz_j)
   \equiv{1\over n'!n!}\int\prod_{i=1}^{n'}d^2\ww'_i
  \prod_{j=1}^{3} |\zz_j-\ww'_i|^{2a'_j}
  \int\prod_{i=1}^{n}d^2\ww_i
  \prod_{j=1}^{3} |\zz_j-\ww_i|^{2a_j}
  \cr&\qquad\times
  \prod_{i<j\atop i,j=1}^{n'}|\ww'_i-\ww'_j|^{4\rt'}
  \prod_{i<j\atop i,j=1}^{n}|\ww_i-\ww_j|^{4\rt}
   \prod_{i=1}^{n'}\prod_{j=1}^{n}\left|\ww'_i-\ww_j\right|^{-2}\cr
   &=\left({\pi\over2}\right)^{n+n'}
   \left(2\rt\right)^{-2nn'-4\frac{{nn'/2}}+n-n'}
   \Delta^{-n'}\!(\rt'+1/2)    \Delta^{-n}\!(\rt+1/2)
   \prod_{i<j\atop1}^3|\zz_i-\zz_j|^{2\delta_{ij}}
   \cr&\qquad\times
\prod_{i=1}^{n'}\Delta(-n/2+\frac{(i+n)/2}+i\rt')
 \prod_{i=0}^{n'-1}\prod_{j=1}^3\Delta(1-n/2-\frac{(i+n)/2}+a'_j+i\rt')
   \cr&\qquad\times
\prod_{i=1}^{n}\Delta(\frac{i/2}+i\rt)
 \prod_{i=0}^{n-1} \prod_{j=1}^{3} \Delta(1-\frac{i/2}+a_j+i\rt)
   \times\cases{1 &$n+n'\in2\IZ$\cr
    \invt\invtb (2\rt')^{-1} & $n,n'-1\in2\IZ$\cr
    \invt\invtb(2\rt)^{-1} & $n-1,n'\in2\IZ$\cr}\cr}
\eqn\Inteq$$
where $\Delta(x)\equiv\Gamma(x)/\Gamma(1-x)$,
$\delta_{12}\equiv n'a'_1+na_1+n'a'_2+na_2-(n'a'_3+na_3)$ and the
other $\delta_{ij}$'s may be obtained using cyclic permutations.
The above formula is valid when
 the integral has no weight with respect to
$\ww_j$, \ie,
$$\sum a'_j  + 2\rt'(n'-1)-(n-1)=0,\qquad
\sum a_j  + 2\rt(n-1)-(n'-1)=0\eqno\eq$$
This condition corresponds just to charge
conservation in the free field theory and guarantees
the covariance of the correlation function under
conformal transformations with respect to $\zz_j$.

Using this integral formula
for $(n',n)=(0,s)$, we find
$$\eqalign{
\int D_{\!\!\hat E}\Phi  \,e^{-S_{SL}}&
\prodone 3 e^{\beta(h_j)\Phi(\zz_j)} =
\mu^s 2^{-3s}\bestar^{-1}{\Gamma(-s)\Gamma(s+1)}
\Delta^{-s}\!(-\rt'+1/2)
\prod_{i<j\atop1}^3|\zz_i-\zz_j|^{-2\gamma_{ij}}
\cr&\quad\times
\!\!\!\prod_{(x',x)=(0,0)\atop (r_j',r_j)}
\prod_{i=1}^{s}\Delta(|2x'\rt'-x|/2+\frac{i/2}-i\rt')
\times\cases{1 &$s$ even\cr \invt\invtb(2\rt')^{-1} &$s$ odd\cr}
\cr}\eqn\Corrgeq$$
We defined
$\rt'\equiv\bestar^2/2$, $\gamma_{12}\equiv h_1+h_2-h_3-1/2$
and so on.
In this formula and in the expression for the correlation function
of the matter fields, there are factors that become ill--defined
as $r_j,r'_j$ approach integers. Below, we shall keep $r_j,r'_j$
to be non--integer and take the limit of $r_j,r'_j$ going to
integers after combining the gravitational and the matter
part of the correlation function. At that point, the expressions
will be well defined.
It is also possible to derive formulas that are well defined
for integer values of $r_j,r'_j$ as in the bosonic case
[\GL,\AD]. However it makes the calculation less transparent
and since it is not necessary, we shall not do so here.

The three point function on the sphere,
$\corr{\prodone 3 \Opcft j}_{\hat E}$, may be computed
using the method analogous to  that of [\KITETAL,\DF].
An arbitrary primary field in the Kac table $\{\Psi^{matter}_{r'r},\
1\leq r'\leq q+2,1\leq r \leq q\}$ may be expressed using
a free field $\varPhi$ as\foot{We normalize
the scalar fields $\Phi,\varPhi$ according to the standard
convention, $\Phi(\zz)\Phi(\zz')=-\ln|\zz-\zz'|^2+\c O(1)$ and
likewise for $\varPhi$.
This differs from the normalization of [\DF] by a factor of
$\sqrt2$.}
$$\Psi^{matter}_{r'r} =e^{i\al{r'r}\varPhi},\quad
\al{r'r}\equiv\half\left[(1-r')\al-+(1-r)\al+\right],\quad
\al-=\bestar=-{1\over\al+}=-\sqrt{q\over q+2} \eqno\eq $$
with the background charge $2\al0$ at infinity
($2\al0\equiv\al++\al-$).
$\varPhi$ is governed by the action
$$S_{\varPhi} ={1\over4\pi} \int{\hat E}
\left[\half D_\alpha\varPhi
D^\alpha\varPhi+i\al0\hat R\varPhi+e^{i\al-\varPhi}+e^{i\al+\varPhi}
\right]\eqno\eq$$
Then the three point function is
$$\eqalign{
\left\langle\prodone 3 \Opcft j\right\rangle_{\!\!\hat E}
\!\!&=\int D_{\!\!\hat E}\varPhi e^{-S_\varPhi}
\prodone3 e^{i\alr j\varPhi(\zz_j)}
{1\over n'!n!}\left(\int e^{i\al-\varPhi}\right)^{n'}
\left(\int e^{i\al+\varPhi}\right)^{n}
\cr&=
\prod_{i<j\atop1}^3|\zz_i-\zz_j|^{2\alr i\alr j}
{\cal J}_{n'n}(\al-\alr j\,,\,\al+\alr j\,,\,\rt,\rt';\zz_j)\cr}
\eqn\Corrcfteq$$
Here,
$$ 2n'\equiv\sum_{j=1}^3r'_j-1,\quad
2n\equiv\sum_{j=1}^3r_j-1,\quad
\rt'=\al-^2/2,\quad\rt=\al+^2/2\eqno\eq$$

We combine the gravitational part and the matter part
of the correlation function, \Corrgeq, \Corrcfteq,
and then integrate over the locations of the operators
$\Op_j$ and divide by the volume of the conformal  automorphisms of
the sphere
to obtain the following expression for the three point function.
$$
\eqalign{\corr{\prodone 3 \Op_j}
&=\mu^s2^{-n'-n-3s}\pi^{n'+n}
\Gamma(-s)\Gamma(s+1)
(2\rt)^{-2nn'+n-n'-4\frac{n/2}+1/2}
\cr&\times
\Delta^{-n'+s}\!(\rt+1/2)\Delta^{-n}\!(\rt+1/2)
\!\!\prod_{(x',x)=(0,0)\atop (r_j',r_j)}\Biggl\{
\prod_{i=1}^{s}\Delta(|2x'\rt'-\rt|/2+\frac{i/2}-i\rt')
\cr&\times
\prod_{i=1}^{n'}\Delta((x-2x'\rt'-n)/2+\frac{(i+n)/2}+i\rt')
\prod_{i=1}^{n}\Delta((x'-2x\rt)/2+\frac{i/2}+i\rt)
\Biggr\}\cr}
  \eqn\Correq$$
Integration over $\invt,\invtb$ requires that
$n'+n+s$ be odd.

We recall that in minimal models, the representation of
primary fields in the Kac table is doubly degenerate,
namely, the fields
$\Psi^{matter}_{r',r}$,  $\Psi^{matter}_{q+2-r',q-r}$
represent the same field.
However, the conformal field theory three point functions
are {\it not} invariant when this interchange is applied to one
of the fields in the correlation function but {\it is}
invariant when the operation is applied to two of the fields.
(Up to normalization factors that may be absorbed into
the normalization of the external fields.)
Since $r_j-2r'_j\rt'$ just changes sign under the operation
$(r',r) \leftrightarrow(q+2-r',q-r)$,
we may assume that $r_j-2r'_j\rt'$
are all of the same sign for the fields in the three point function
without any loss in generality.

When $r_j-2r'_j\rt' \geq0$ for $j=1,2,3$, from \Dimeq\ and \Seq, we obtain
$$2\rt'={n\over n'+s+1}\eqno\eq$$
As shown in the Appendix, the following formula holds when
$n+n'+s$ is odd.
$$ \eqalign{\prod_{i=1}^s\Delta & (y/2+\frac{i/2}-i\rt')  =
   (2\rt)^{(n+1-2y)(n'+s+1)/2-n/2+\frac{n/2}}\Delta^{-1}\!(y/2)
\cr&\qquad\times
\prod_{i=1}^n\Delta^{-1}\!(\frac{i/2}+(-y+i)\rt)
\prod_{i=1}^{n'}   \Delta^{-1}\!((y-n)/2+\frac{(i+n)/2}+i\rt')\cr}
\eqn\Rearr$$
Using this formula for $y=0,r_j-2r'_j\rt'$ and
from \Correq, we obtain the formula for the three point function as
$$\eqalign{
\corr{\prodone 3 \Op_j}
&=\mu^s 2^{-n'-n-3s}\pi^{n'+n}
(2\rt)^{s+3/2}\Delta^{-n'+s}(\rt'+1/2)
\Delta^{-n}(\rt+1/2)
\cr&\qquad\times
\prodone3 \Delta^{-1}\!((r_j-2r'_j\rt')/2)\cr} \eqn\Corrfeq$$
A subtlety needs to be mentioned here; as in the bosonic case [\GL],
the expression here contains a factor $\Gamma(-s)\Gamma(s+1)S(s)$
which is of indeterminate form $0/0$ when $s$ is integer.
This factor is taken to be
one.  This formal procedure is justified by the
semi--classical analysis which establishes the identity as
$s$ goes to infinity in the complex $s$ plane [\AD].

Rescaling  the external fields as
$$\eqalign{
\Op_j&\mapsto const.\times2^{-2r'_j+(-1/2+3\rt)r_j}\pi^{(r'_j+r_j)/2}
(2\rt)^{\rt r_j-r'_j/2}
\Delta^{\rt r_j-r'_j}\!(\rt'+1/2)
\cr&\qquad\times
\Delta^{-r_j/2}\!(\rt+1/2)
\Delta^{-1}\!((r_j-2r'_j\rt')/2)\Op_j\cr}\eqn\rescaling$$
the three point function reduces just to
$$\corr{\prodone 3 \Op_j} =\mu^s\eqno\eq$$
The constant factor in the formula \rescaling\
denotes  a factor that does not depend on the external indices
$\{(r'_j,r_j)\}$.

When $r_j-2r'_j\rt' \leq0$ for $j=1,2,3$,  we analogously obtain
$$2\rt'={n+1\over n'-s}\eqno\eq$$
As shown in the Appendix, the following formula holds when
$n+n'+s$ is odd.
$$ \eqalign{\prod_{i=1}^s&\Delta(-y/2+\frac{i/2}-i\rt')  =
   (2\rt)^{(n-2y)(n'-s)/2-n/2+\frac{n/2}-1}\Delta^{-1}\!(-y\rt)
\cr&\qquad\times
\prod_{i=1}^n\Delta^{-1}\!(\frac{i/2}+(-y+i)\rt)
\prod_{i=1}^{n'}   \Delta^{-1}\!((y-n)/2+\frac{(i+n)/2}+i\rt')\cr}
\eqn\Rearrinv$$
Again, using this formula for $y=0,r_j-2r'_j\rt'$ we obtain
the formula for the correlation function.
The result is identical to \Corrfeq, except for the replacement
$$ \Delta^{-1}\!((r_j-2r'_j\rt')/2)\mapsto
\Delta^{-1}\!((r'_j-2r_j\rt)/2).\eqno\eq$$
Therefore, the correlation function again reduces to $\mu^s$ after
rescaling the fields.

Differentiating with respect to $\mu$ is equivalent to
bringing down the area operator:
$${\d\over\d\mu}\corr{\prodone N\Op_j}
=\corr{{\bf 1}\prodone N\Op_j}\eqno\eq$$
The three point function which is independent of the normalization
of the fields may be computed using this relation as
$${\corr{\Op_1\Op_2\Op_3}^2{\rm Z}\over
\corr{\Op_1\Op_1}\corr{\Op_2\Op_2}\corr{\Op_3\Op_3}}
 ={\prodone3\left|qr'_j-(q+2)r_j\right|\over4(q+1)(q+2)}\eqno\eq$$
where ${\rm Z}$ is the partition function of the model on the sphere.
\chapter{Summary and Discussion}
In this paper, we computed the three point functions on the sphere
of Neveu--Schwarz primary fields in $N=1$
minimal matter coupled to supergravity.
The result is simple; the three point function reduces
to just the power of the chemical potential for the area raised
to the power of the scaling dimensions.
This general feature
is consistent with the expectations from the matrix models and topological
models.
However, the corresponding (super--)matrix models are yet to be
identified, despite some previous effort [\SUPERDDK,\AG].
The computations of correlation functions put
quantum supergravity on a more concrete ground,  and may
provide the basis for identifying the corresponding matrix
and topological models.

It should be possible to build higher genus, higher point
correlation functions by sewing together genus zero
three point functions.  Therefore, specifying the spectrum
of the model uniquely identifies the model.
This important point
may be established by computing higher genus or higher point
functions.
In conformal field theory, modular invariance strongly restricts
the spectrum, and in the case of minimal models, the classification
is known [\CIZ,\CAPPELI].
It is natural to assume that the spectrum of the conformal field
theory model is unchanged when coupled to gravity or supergravity.
This is also consistent with the fact that the three
point functions between general primary fields in the Kac
table reduces to the chemical potential raised to the power of
the sum of the scaling dimensions.
Such is the case in the left--right symmetric (``A--type")
bosonic unitary minimal models coupled to
gravity and the corresponding matrix models in this case
are Hermitean matrix models.
In the non--unitary models, even in the bosonic case, the
situation is not clear and it is of interest to find the complete
spectrum [\BD].

For $N=1$ minimal models with even $q$, the scaling dimensions
of the Neveu--Schwarz fields reduce to that of the bosonic minimal
model if one restricts the indices of the primary fields
to lie in about a quarter of the Kac table.
It has even been  suggested that the supersymmetric model
reduces to the bosonic model where such a truncation of the spectrum
occurs [\SUPERDDK].
This is possible, but unexpected from our results and also
seems to disagree with the one--loop partition function [\BK].
Also, one might expect that there is a qualitative difference
for $q$ even and $q$ odd, since the Witten index is zero when $q$ is odd
[\SUPERDDK]. We find no qualitative difference
between $q$ even and odd at tree level.
Perhaps a promising candidate for  the corresponding matrix models
are unitary matrix models [\PS].
They have the feature that the scaling dimensions of some of the
operators agree with the Hermitean matrix models, present in
$N=1$ minimal matter.

Even though coupling conformal matter to Liouville theory seems
like the  natural  approach from field theory, the computations
needed were involved and yet the result simple.
It would be interesting to find a method of computation in this
or in another
approach which obtains the same result in a more straightforward way.
If no such scheme is found, Liouville theory might not be
the appropriate framework for computing correlation functions
of matter coupled to gravity.
However, the super--Liouville approach offers the most
straightforward generalization from the bosonic case and
indeed we note that correlation functions in matter coupled to
two dimensional
supergravity have not been computed
in the other approaches.
An important point still to be investigated is how to incorporate
 Ramond fields, since they  break two dimensional supersymmetry.
\medskip
\noindent\undertext{Acknowledgments}\nl
One of us (E.D) acknowledges  useful discussions with
\v Cedomir Crnkovi\'c and Vipul Periwal
and the hospitality of the Aspen Center for Physics
where part of this work was carried out.
\appendix
In the paper, the identities, \Rearr\  and \Rearrinv\  were used.
Since the proof is essentially the same for both cases,
we shall only present the case \Rearrinv,  when
$2\rt'={(n+1)/(n'-s)}$ and $n'+n+s$ is odd.

Call the left hand side of \Rearrinv\ $L(y)$ and right hand side $R(y)$.
We shall proceed in two steps;
\item{(i)} $L(y)/R(y)$ is periodic in $y$ with period $1-2\rt'$.
\item{(i\!i)} $L(y)/R(y)$  goes to 1 as $y$ goes to infinity in
the complex  $y$ plane.
\par
{}From
$$\eqalign{
{ L(y) \over L(y+1-2\rt') } &= (2\rt)^{-s+2\frac{s/2}}\!
{\Delta(-y/2+\frac{s/2}-s\rt')\Delta\left((y+1)\rt+(s+1)/2-\frac{s/2}\right)
\over
\Delta(-y/2)\Delta((y+1)\rt+1/2)}\cr
{R(y+1-2\rt') \over R(y)} &= (2\rt)^{s+1-2\frac{n/2}-2\frac{n'/2}}
{\Delta(-y/2)\Delta\left(1-\frac{n/2}+(y-n)\rt\right)
\over
\Delta\left((y-n)/2+1-\frac{n/2}\right)\Delta\left(1/2-(y+1)\rt\right)}
\cr&\qquad\times
{\Delta\left((-y+n)\rt+\frac{n/2}\right)
\Delta\left((y-n)/2+1-\frac{n/2}-\frac{n'/2}+n'\rt'\right)
\over
\Delta\left((-y+n)/2+\frac{n/2}\right)
\Delta\left((y-n)\rt+n'/2+1-\frac{n/2}-\frac{n'/2}\right)}
\cr}\eqno\eq$$
Multiplying one formula by the other,
it is easy to establish the statement $(i)$ when $n'+n+s$ is odd.
To show $(i\!i)$,  we use Stirling's formula,
$$\ln\Delta(z)=(z-\half)\ln(-e^2z^2)-1+\c O(\inv z)\eqno\eq$$
Then by a straightforward, but tedious calculation, we
see that
$$\ln L(y)/R(y)=\left(\frac{s/2}+\frac{n/2}+\frac{n'/2}
-4\frac{n/2}\frac{n'/2}
-1/2\right)\ln(ie^2y/2)+\c O(\inv y)\eqno\eq$$
The right hand side of the equation vanishes when $n'+n+s$ is
odd, so that we show statement $(i\!i)$ in this case.
This in turn completes the derivation of the formula
\Rearrinv. The proof of \Rearr\ is essentially the same.

When $n$ is odd, \Rearrinv\  may also be shown by factoring
out two from the numerator and the denominator of $\rt'$
and using the formulas derived in the bosonic case [\GL,\AD].
Likewise, when $n$ even in \Rearr, the formula may again be
reduced to the bosonic case.
\refout
\endpage\end

{}From the field theory point of view of (super--)Liouville
model coupled to matter, it is not clear why this is the
case and indeed, the computations involved are non--trivial.